\documentclass{emulateapj}

\usepackage{natbib}
\usepackage{amsmath}

\newcommand{\gri}{\protect\hbox{$gri$} }
\newcommand{\griz}{\protect\hbox{$griz$} }

\newcommand{\vri}{\protect\hbox{$V\!RI$} }

\newcommand{\about}{$\sim\!\!$~}
\newcommand{\kms}{\,km\,s$^{-1}$}

\def\lsim{\hbox{\rlap{\raise 0.425ex\hbox{$<$}}\lower 0.65ex\hbox{$\sim$}}}
\def\gsim{\hbox{\rlap{\raise 0.425ex\hbox{$>$}}\lower 0.65ex\hbox{$\sim$}}}

\newcommand{\sn}{SN~2009ku}

\newcommand{\snno}{2009ku}

\shorttitle{Observations of \sn}
\shortauthors{Narayan et~al.}

\begin{document}

 \title{Displaying the Heterogeneity of the SN~2002cx-like Subclass of
 Type~Ia Supernovae with Observations of the Pan-STARRS-1 Discovered
 \sn\altaffilmark{*}}

\def\harv{1}
\def\cfa{2}
\def\clay{3}
\def\stsci{4}
\def\jhu{5}
\def\qub{6}
\def\ifa{7}
\def\prin{8}
\def\usno{9}
\def\pitt{10}

\author{
{G.~Narayan}\altaffilmark{\harv},
{R.~J.~Foley}\altaffilmark{\cfa,\clay}, 
{E.~Berger}\altaffilmark{\cfa},
{M.~T.~Botticella}\altaffilmark{\qub},
{R.~Chornock}\altaffilmark{\cfa},
{M.~E.~Huber}\altaffilmark{\jhu},
{A.~Rest}\altaffilmark{\cfa,\stsci},
{D.~Scolnic}\altaffilmark{\jhu},
{S.~Smartt}\altaffilmark{\qub},
{S.~Valenti}\altaffilmark{\qub},
{A.~M.~Soderberg}\altaffilmark{\cfa},
{W.~S.~Burgett}\altaffilmark{\ifa},
{K.~C.~Chambers}\altaffilmark{\ifa},
{G.~Gates}\altaffilmark{\ifa},
{T.~Grav}\altaffilmark{\jhu},
{N.~Kaiser}\altaffilmark{\ifa},
{R.~P.~Kirshner}\altaffilmark{\cfa},
{E.~A.~Magnier}\altaffilmark{\ifa},
{J~.S.Morgan}\altaffilmark{\ifa},
{P.~A.~Price}\altaffilmark{\ifa,\prin},
{A.~G.~Riess}\altaffilmark{\jhu,\stsci}
{C.~W.~Stubbs}\altaffilmark{\harv},
{W.~E.~Sweeney}\altaffilmark{\ifa},
{J.~L.~Tonry}\altaffilmark{\ifa},
{R.~J.~Wainscoat}\altaffilmark{\ifa}
{W.~M.~Wood-Vasey}\altaffilmark{\pitt}
}

\altaffiltext{*}{This paper includes data gathered with the 6.5~m
Magellan Telescopes located at the Las Campanas Observatory, Chile;
the Subaru Telescope, which is operated by the National Astronomical
Observatory of Japan; the Nordic Optical Telescope, operated by
Denmark, Finland, Iceland, Norway, and Sweden and the Liverpool
Telescope operated by the Liverpool John Moores University with
financial support from the UK Science and Technology Facilities
Council on the island of La Palma in the Spanish Observatorio del
Roque de los Muchachos of the Instituto de Astrofisica de Canarias;
and the Gemini Observatory, which is operated by the Association of
Universities for Research in Astronomy, Inc., under a cooperative
agreement with the NSF}

\altaffiltext{\harv}{
Department of Physics,
Harvard University,
17 Oxford Street,
Cambridge, MA 02138
}
\altaffiltext{\cfa}{
Harvard-Smithsonian Center for Astrophysics,
60 Garden Street, 
Cambridge, MA 02138
}
\altaffiltext{\clay}{
Clay Fellow. Electronic address rfoley@cfa.harvard.edu .
}
\altaffiltext{\stsci}{
Space Telescope Science Institute,
3700 San Martin Drive,
Baltimore, MD 21218
}
\altaffiltext{\jhu}{
Department of Physics and Astronomy,
Johns Hopkins University,
3400 N. Charles Street, 
Baltimore, MD 21218
}
\altaffiltext{\qub}{
Astrophysics Research Centre,
School of Mathematics and Physics,
Queen's University Belfast,
Belfast, BT7 1NN, UK
}
\altaffiltext{\ifa}{
Institute for Astronomy,
University of Hawaii at Manoa,
Honolulu, HI 96822
}
\altaffiltext{\prin}{
Department of Astrophysical Sciences,
Princeton University,
Princeton, NJ 08544
}
\altaffiltext{\pitt}{
Department of Physics and Astronomy,
University of Pittsburgh,
3941 O'Hara Street,
Pittsburgh PA 15260
}
\email{gnarayan@cfa.harvard.edu}

\begin{abstract}
\sn, discovered by Pan-STARRS-1, is a Type Ia supernova (SN~Ia), and a
member of the distinct SN~2002cx-like class of SNe~Ia.  Its light
curves are similar to the prototypical SN~2002cx, but are slightly
broader and have a later rise to maximum in $g$.  \sn\ is brighter
(\about 0.6~mag) than other SN~2002cx-like objects, peaking at $M_{V}
= -18.4$~mag --- which is still significantly fainter than typical
SNe~Ia.  \sn, which had an ejecta velocity of \about 2000~\kms\ at
18~days after maximum brightness is spectroscopically most similar to
SN~2008ha, which also had extremely low-velocity ejecta.  However,
SN~2008ha had an exceedingly low luminosity, peaking at $M_{V} =
-14.2$~mag, \about 4~mag fainter than \sn.  The contrast of high
luminosity and low ejecta velocity for \sn\ is contrary to an emerging
trend seen for the SN~2002cx class.  \sn\ is a counter-example of a
previously held belief that the class was more homogeneous than
typical SNe~Ia, indicating that the class has a diverse progenitor
population and/or complicated explosion physics.  As the first example
of a member of this class of objects from the new generation of
transient surveys, \sn\ is an indication of the potential for these
surveys to find rare and interesting objects.
\end{abstract}

\keywords{supernovae: general --- supernovae: individual(\sn)}


\section{Introduction}\label{s:intro}

Most Type Ia supernovae (SNe~Ia) can be described by a single
parameter that relates peak luminosity with light-curve shape
\citep{Phillips93}, intrinsic color \citep{Riess96}, and $^{56}$Ni
mass \citep{Mazzali07}.  There are examples of particular SNe~Ia not
following this parameterization \citep[e.g.,][]{Li01:00cx,
Foley10:06bt}, with a single, relatively large subclass of objects
(see \citealt{Foley09:08ha} for a recent list of members) dominating
the outliers.  Members of this subclass, labeled ``SN~2002cx-like''
after the prototypical object \citep{Li03:02cx}, have peak magnitudes
\about 2~mag below that of normal SNe~Ia and spectra that resemble the
high-luminosity SN~Ia 1991T \citep{Filippenko92:91T, Phillips92} at
early and intermediate phases, except with significantly lower
expansion velocities and having late-time (\about 1 yr after maximum)
spectra which show low-velocity \ion{Fe}{2} lines, few lines from
intermediate-mass elements, and no strong forbidden lines
\citep{Jha06:02cx, Sahu08}.  The extreme characteristics of this class
can be explained if the objects are full deflagrations of a white
dwarf (WD) \citep{Branch04, Phillips07}.  Because of their low
velocities, which eases line identification, and probing of the
deflagration process, which is essential to all SN~Ia explosions, this
subclass is particularly useful for understanding typical SN~Ia
explosions.  For a review of this class, see
\citet{Jha06:02cx}.

A recent addition to this class, SN~2008ha \citep{Foley09:08ha,
Valenti09, Foley10:08ha}, was much fainter (peaking at $M_{V} =
-14.2$~mag) and had a significantly lower velocity ($v \approx
2000$~\kms) than the typical member.  Although its maximum-light
spectrum indicates that the object underwent C/O burning
\citep{Foley10:08ha}, certain observations are consistent with a
massive-star progenitor \citep{Foley09:08ha, Valenti09,
Moriya10}. SN~2008ha generated \about $10^{-3} M_{\sun}$ of $^{56}$Ni
and ejected \about $0.3 M_{\sun}$ of material \citep{Foley10:08ha},
suggesting that the most plausible explanation was a failed
deflagration of a WD that did not destroy the progenitor star
\citep{Foley09:08ha, Foley10:08ha}.  Furthermore, another member of
this class, SN~2008ge, was hosted in an S0 galaxy with no signs of
star formation or massive stars, including at the SN position in
pre-explosion {\it HST} imaging
\citep{Foley10:08ge}.

Using a small sample of SN~2002cx-like objects, \citet{McClelland10}
suggested that their peak luminosities correlated with light-curve
shapes and ejecta velocity (with more luminous SNe having faster
ejecta and slower declining light curves).  These correlations are
similar to that of the relationship between light-curve shape and peak
luminosity in ``normal'' SNe~Ia \citep{Phillips93} and suggest that
SN~2002cx-like objects can possibly be described by a single
parameter.

As part of the Medium Deep Survey (MDS) \sn\ was discovered by
Pan-STARRS-1 (PS1) on 2009 Oct.\ 4.83 (UT dates are used throughout
this paper) at mag 19.9 \citep{Rest09} in APMUKS(BJ)
B032747.73-281526.1, an Sc galaxy with a redshift of 0.0792 in the
MD02 field (coincident with the CDF-S field).  Spectroscopy showed
that it was a SN~Ia similar to SN~2002cx \citep{Rest09}.  From the
discovery magnitude and redshift alone, \sn\ had a peak magnitude of
$M_{r} \lesssim -18$~mag, similar to that of SN~2002cx, but much
brighter than SN~2008ha.  With our full light curves, we measure its
peak brightness below.  We also show that the spectra of \sn\ are
similar to SN~2002cx and most similar to those of SN~2008ha, with both
objects having small expansion velocities.

We present and discuss our observations in Section~\ref{s:obs}.  In
Section~\ref{s:analysis}, we place \sn\ in the context of other
objects in the SN~2002cx-like subclass and typical SNe~Ia.  We discuss
our results and summarize our conclusions in Section~\ref{s:disc}.
Throughout this work, we use the concordance cosmology of ($H_{0}$,
$\Omega_{m}$, $\Omega_{\Lambda}) = (70$, 0.3, 0.7).


\section{Observations and Data Reduction}\label{s:obs}

\sn\ was discovered before maximum brightness by the 1.8~m PS1
telescope on Haleakala as part of the rolling MDS search.
Unfortunately, PS1 coverage of \sn\ was interrupted on 1 October 2009,
shortly after maximum brightness, for engineering work on the
telescope.

To supplement the PS1 light curve, we obtained \gri\ photometry with
GMOS on the Gemini-South 8~m telescope \citep{Hook04}, RATCam on the
Liverpool 2~m telescope, and Suprime-Cam on the Subaru 8.2~m
telescope.  Late-time \griz\!\! photometry was resumed with PS1 beginning
15 December 2009.

The images were de-biased, flat-fielded, and astrometrically registered
by pipelines developed specifically for each imager.  PS1 Gigapixel
Camera (GPC1) data was processed with the Pan-STARRS Image Processing
Pipeline\footnote{http://svn.pan-starrs.ifa.hawaii.edu/trac/ipp/wiki/
.}. GMOS data was reduced using the Gemini IRAF\footnote{IRAF: the
Image Reduction and Analysis Facility is distributed by the National
Optical Astronomy Observatory, which is operated by the Association of
Universities for Research in Astronomy, Inc. (AURA) under cooperative
agreement with the National Science Foundation (NSF).} package.
RATCam and Suprime-Cam data was processed by a modified version of the
SuperMacho pipeline \citep{Rest05}.

Following preliminary processing, we used the SuperMacho pipeline for
photometry and difference imaging. Photometric catalogs of secondary
standard stars were produced by first matching the GPC1 MDS fields
03--07 to images obtained by the Sloan Digital Sky Survey (SDSS) and
then extrapolating the zero points to observations of the MD02 on the
same night.

We performed template subtraction to remove the underlying host-galaxy
light and measure the SN flux.  A set of reference \griz\!\! images
were constructed from deep stacks of GPC1 images obtained in
photometric conditions in late January 2010.  This procedure allows a
reference set to be constructed from homogeneous imaging while being
sufficiently deep for difference imaging with telescopes of much
larger aperture.

HOTPANTS\footnote{http://www.astro.washington.edu/users/becker/hotpants.html
.}  was used to determine a convolution kernel and subtract this
reference image from each image.  A modified version of DoPHOT
\citep{Schechter93} was then used to measure the flux of the SN.

Comparing our reference images to deep stacks produced from the
last available PS1 epoch, we found low-level residual flux at the
position of the SN.  As this flux is not detected in single images and
at this epoch \sn\ is more than 80~days past maximum brightness, we do
not believe that this residual flux significantly affects the
photometry.  Nonetheless, an offset is added to all photometry to
account for the residual flux.

We present the \griz light curves in Figure~\ref{f:lc} and
Table~\ref{t:phot}.

\begin{figure}
\begin{center}
\epsscale{1.3}
\rotatebox{90}{
\plotone{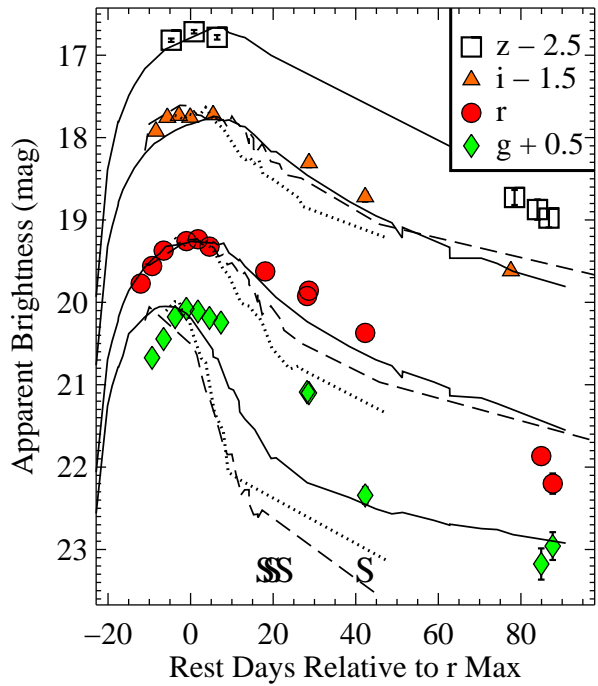}}
\caption{\griz light curves of \sn.  The uncertainties for
most data points are smaller than the plotted symbols. Also plotted
are comparison light curves of SNe~2002cx (\vri\!\!; dashed lines),
2005hk (\griz\!\!; solid lines), and 2008ha (\vri\!\!; dotted lines)
after applying a magnitude offset to match the peak in each band.  The
epoch of each spectrum is marked with an `S.'}\label{f:lc}
\end{center}
\end{figure}

\begin{deluxetable}{rccc}
\tabletypesize{\scriptsize}
\tablewidth{0pt}
\tablecaption{Photometric Observations\label{t:phot}}
\tablehead{
\colhead{Julian Date} &
\colhead{} &
\colhead{} &
\colhead{} \\
\colhead{$-2455000$}&
\colhead{Passband} &
\colhead{Magnitude\tablenotemark{a}} &
\colhead{Telescope}}
\startdata
 84.12  & $r$ &  19.77(03) &     PS1 \\
 87.10  & $g$ &  20.17(03) &     PS1 \\
 87.12  & $r$ &  19.56(02) &     PS1 \\
 88.10  & $i$ &  19.42(02) &     PS1 \\
 90.11  & $g$ &  19.94(02) &     PS1 \\
 90.13  & $r$ &  19.37(02) &     PS1 \\
 91.09  & $i$ &  19.26(01) &     PS1 \\
 92.11  & $z$ &  19.32(02) &     PS1 \\
 93.11  & $g$ &  19.68(02) &     PS1 \\
 94.11  & $i$ &  19.23(02) &     PS1 \\
 96.07  & $g$ &  19.58(02) &     PS1 \\
 96.09  & $r$ &  19.26(01) &     PS1 \\
 97.06  & $i$ &  19.25(01) &     PS1 \\
 98.09  & $z$ &  19.21(02) &     PS1 \\
 99.08  & $g$ &  19.61(02) &     PS1 \\
 99.09  & $r$ &  19.23(02) &     PS1 \\
102.10  & $g$ &  19.69(02) &     PS1 \\
102.11  & $r$ &  19.32(02) &     PS1 \\
103.07  & $i$ &  19.22(02) &     PS1 \\
104.12  & $z$ &  19.28(03) &     PS1 \\
105.10  & $g$ &  19.74(02) &     PS1 \\
116.68  & $r$ &  19.62(02) &     Gemini-S \\
127.62  & $g$ &  20.59(05) &     Liverpool \\
127.62  & $r$ &  19.92(04) &     Liverpool \\
128.08  & $r$ &  19.86(06) &     Subaru \\
128.10  & $g$ &  20.60(06) &     Subaru \\
128.11  & $i$ &  19.81(06) &     Subaru \\
142.81  & $g$ &  21.84(11) &     Gemini-S \\
142.81  & $r$ &  20.37(03) &     Gemini-S \\
142.81  & $i$ &  20.22(06) &     Gemini-S \\
180.90  & $i$ &  21.12(03) &     PS1 \\
181.85  & $z$ &  21.23(10) &     PS1 \\
187.84  & $z$ &  21.37(13) &     PS1 \\
188.82  & $g$ &  22.68(21) &     PS1 \\
188.83  & $r$ &  21.87(10) &     PS1 \\
190.83  & $z$ &  21.48(12) &     PS1 \\
191.82  & $g$ &  22.46(18) &     PS1 \\
191.83  & $r$ &  22.20(13) &     PS1 \\
213.79  & $i$ &  22.85(76) &     PS1 \\
214.80  & $z$ &  22.99(63) &     PS1 \\
219.76  & $i$ &  23.86(86) &     PS1 \\
220.76  & $z$ &  22.98(68) &     PS1
\enddata
\tablenotetext{a}{Uncertainties are reported in hundredths of a magnitude}
\end{deluxetable}

We obtained two low-resolution spectra of \sn\ with GMOS
\citep{Hook04} on Gemini-South (PI Berger; Program GS-2009B-Q-30) and
one with the ALFOSC
spectrograph\footnote{http://www.not.iac.es/instruments/alfosc .} on
the 2.5~m Nordic Optical Telescope.  We also obtained a
medium-resolution spectrum of \sn\ with the MagE spectrograph
\citep{Marshall08} on the Magellan Clay 6.5~m telescope.  A journal of
observations can be found in Table~\ref{t:spec}.  Standard CCD
processing and spectrum extraction were performed with IRAF.  The data
were extracted using the optimal algorithm of \citet{Horne86}.
Low-order polynomial fits to calibration-lamp spectra were used to
establish the wavelength scale, and small adjustments derived from
night-sky lines in the object frames were applied.  For the MagE
spectrum, the sky was subtracted from the images using the method
described by \citet{Kelson03}.  We employed our own IDL routines for
flux calibration and telluric line removal (except for the ALFOSC
spectrum) using the well-exposed continua of spectrophotometric
standard stars \citep{Wade88, Foley03, Foley06, Foley09:08ha}.  Our
spectra of \sn\ are presented in Figure~\ref{f:spec}.

\begin{deluxetable}{l@{ }l@{ }l@{ }r@{ }l}
\tabletypesize{\scriptsize}
\tablewidth{0pt}
\tablecaption{Log of Spectral Observations\label{t:spec}}
\tablehead{
\colhead{} &
\colhead{} &
\colhead{Telescope /} &
\colhead{Exposure} &
\colhead{} \\
\colhead{Phase\tablenotemark{a}} &
\colhead{UT Date} &
\colhead{Instrument} &
\colhead{(s)} &
\colhead{Observer\tablenotemark{b}}}

\startdata

18.1 & 2009 Oct.\ 12.2 & Gemini-S/GMOS & $2 \times 1200$ & EC \\
19.9 & 2009 Oct.\ 14.1 & NOT/ALFOSC    & 3600            & SM, EK \\
22.8 & 2009 Oct.\ 17.2 & Magellan/MagE & $3 \times 1800$ & RF \\
42.3 & 2009 Nov.\  7.3 & Gemini-S/GMOS & $2 \times 1200$ & LF

\enddata

\tablenotetext{a}{Rest-frame days since $r$ maximum, 2009 Sept.\ 22.7 (JD 2,455,097.2).}

\tablenotetext{b}{EC = E.\ Christensen; LF = L.\ Fuhrman; EK = E.\
Kankare; RF = R.\ Foley; SM = S.\ Mattila}

\end{deluxetable}

\begin{figure*}
\begin{center}
\epsscale{1.15}
\plotone{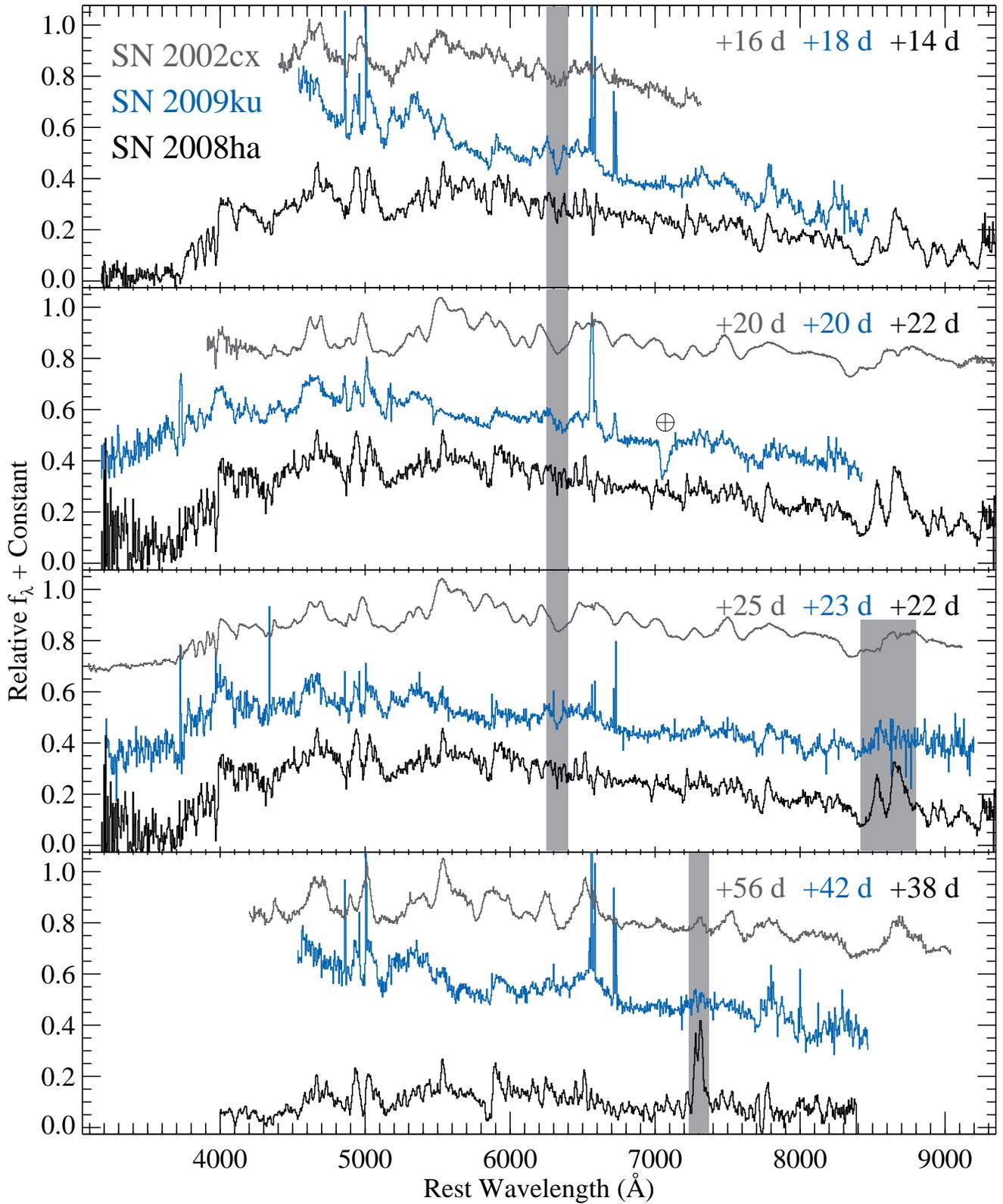}
\caption{Optical spectra of \sn\ (blue curves in the middle of each
panel) denoted by their phase relative to maximum brightness.  The
grey and black curves (top and bottom in each panel) are of spectra of
SNe~2002cx and 2008ha.  For visual comparison, an Sc template galaxy
spectrum has been added to each SN~2002cx and SN~2008ha spectrum.  The
regions corresponding to \ion{Si}{2} $\lambda 6355$, [\ion{Ca}{2}]
$\lambda\lambda 7291$, 7324, and the \ion{Ca}{2} NIR triplet are
shaded grey.  Incomplete telluric correction in the second spectrum is
marked.}\label{f:spec}
\end{center}
\end{figure*}


\section{Analysis}\label{s:analysis}

\subsection{Photometric Analysis}\label{ss:phot}

Figure~\ref{f:lc} shows the \griz light curves of \sn.  Also plotted
are the \vri light curves of SNe~2002cx and 2008ha and the \griz\
light curves of SN~2005hk, a well-observed SN extremely similar to
SN~2002cx \citep{Phillips07}.  Each light curve has been shifted to
have $r$ (or $R$) maximum correspond to $t = 0$~days and so that each
comparison curve peaks at the same magnitude as \sn.

Although its light curve is sparse after maximum, relative to
SNe~2002cx, 2005hk, and 2008ha, \sn\ clearly peaked significantly
later in $g$ than in $r$ and $i$ and had a slower decline in all
bands.  Although the light curves are sparse, we measure $\Delta
m_{15} = 0.53 \pm 0.27$, $0.32 \pm 0.05$, and $0.18 \pm 0.09$~mag in
\gri\!\!, respectively.  The uncertainty in these measurements is
relatively large due to the gap between measurements after maximum
light.  As long as the the light curves are smooth during this period,
our measurements of $\Delta m_{15}$ should not be significantly
affected.  In particular, at $t = 17.9$~days, the $r$ band had
declined by 0.37~mag from peak, which is a strict upper limit on
$\Delta m_{15}(r)$.

Since \sn\ was not observed in $B$ and most comparison objects were
not observed in $g$, we have focused on the $r$/$R$ observations, in
which we expect objects to have a similar light-curve shapes.
However, using the relationships between the decline rates of \gri and
$B$ for SN~2005hk and scaling the decline rates for \sn, we have
determined that \sn\ had $0.51 \le \Delta m_{15} (B) \le 0.67$~mag,
with an average value of 0.59~mag.  This is an extremely small value
of $\Delta m_{15} (B)$ for any SN~Ia.  Although \sn\ has a slower
decline than other SN~2002cx-like objects in \gri\!\!, the uncertainty
in converting from decline rates in \gri to $B$ and the sparsely
sampled post-maximum light curves of \sn\ make us question the
accuracy of the derived value of $\Delta m_{15} (B)$.

\subsection{Spectral Analysis}\label{ss:spec}

Despite contamination by host-galaxy light, all spectra of \sn\
exhibit clear SN features.  These features are at low velocity, and
there are no obvious hydrogen or helium features (except from narrow
galactic emission lines).

The comparison of the spectra of \sn\ to those of other SNe is aided
by the addition of galaxy template spectra to the comparison SN
spectra.  In Figure~\ref{f:spec}, we compare \sn\ to SNe~2002cx and
2008ha at similar phases.  Although the features in the \sn\ spectra
are relatively weak, they match those of SN~2008ha to a high degree.
There is some similarity to SN~2002cx, but also clear differences.
Some of these differences are the result of different line velocities
and widths, while others are likely differences in the composition
and/or opacity of the ejecta.  By comparing \sn\ to SNe~2002cx and
2008ha, it is clear that there are several features in the spectra of
\sn\ corresponding to \ion{Cr}{2}, \ion{Fe}{2}, and \ion{Co}{2} (see
\citealt{Foley09:08ha} for a synthetic spectrum identifying these
features).  Additionally, the spectra of SNe~2008ha and \snno\ evolve
similarly.  From the spectral comparisons, we determine that the
ejecta velocity of \sn\ is similar to that of SN~2008ha, although
perhaps slightly lower (i.e., the \ion{O}{1} $\lambda 7774$ line is at
a lower velocity at similar epochs in \sn); \sn\ must have had a very
low ejecta velocity near maximum.

Examining the spectra in detail, there are a few obvious differences
between SNe~2008ha and \snno.  \sn\ has a clear feature at \about
6325~\AA\ until at least 23~days after maximum brightness.  We
interpret this feature as \ion{Si}{2} $\lambda 6355$, the hallmark of
SNe~Ia, blueshifted by \about 1400~\kms.  This feature is present in
SN~2008ha at maximum brightness \citep{Foley10:08ha}, but is not
clearly present a week later \citep{Foley09:08ha}.  This feature is
present, but weak, in SN~2002cx until around 15~days after maximum
brightness \citep{Li03:02cx, Branch04}.  For SN~2002cx, a feature at a
similar wavelength persists until at least 25~days after maximum
brightness, but this has been interpreted as \ion{Fe}{2}
\citep{Branch04}.

Additionally, SN~2008ha has strong emission from the \ion{Ca}{2} NIR
triplet at 23~days after maximum brightness, but this feature is much
weaker in \sn\ at a similar epoch, being very similar to that of
SN~2002cx.  At \about 40~days after maximum brightness, SN~2008ha has
strong [\ion{Ca}{2}] $\lambda\lambda 7291$, 7324 emission, while
SNe~2002cx and \snno\ have weak or absent [\ion{Ca}{2}] emission at
this epoch.  These differences may indicate that \sn\ burned more
material to Si and less to Ca than SN~2008ha, making its
nucleosynthesis more similar to SN~2002cx.

\subsection{Comparison to Other Objects}

Although the number of SN~2002cx-like objects has grown such that
properties of the class can be examined, few comprehensive studies
have been performed.  \citet{Jha06:02cx} was the first to compile
several members, showing that the relatively small sample of objects
had ``striking spectral homogeneity.''  SN~2008ha, if a true member of
this class, is an outlier in this regard (e.g., the maximum-light
spectrum of SN~2008ha shows strong \ion{Si}{2}, which the other
members do not; \citealt{Foley10:08ha}).  \citet{McClelland10}
suggested (using SNe~2002cx, 2005hk, 2007qd, and 2008ha) that there
were correlations between peak luminosity, light-curve shape and
ejecta velocity.

\sn\ challenges that assertion.  In Figure~\ref{f:prop}, we plot
$\Delta m_{15} (R)$ (or $\Delta m_{15} (r)$ for SNe~2007qd and
\snno), ejecta velocity \about 10~days after maximum, and $M_{V}$ at
peak for several SN~2002cx-like objects and typical SNe~Ia.  We assume
that since SNe~2008ha and \snno\ have a similar ejecta velocity at
18~days after maximum brightness that they have a similar velocity a
week earlier.  Using SN~2005hk, the light curves of \sn\ were
transformed to determine $M_{V}$ at peak.  We assume that there is
negligible host-galaxy extinction, which is supported by the
relatively blue colors of \sn.  Regardless, \sn\ must have been at
least as bright as our value of $M_{V} = -18.4$~mag at peak.  There is
no clear correlation between $\Delta m_{15} (R)$ and $M_{V}$ for
SN~2002cx-like objects, even when restricting to the sample of
\citet{McClelland10}.  The five objects with published values
(including \sn) have a large range in $M_{V}$.  Although most
SN~2002cx-like objects have similar decline rates in $R$, SN~2008ha is
significantly faster than the other objects.

\begin{figure}
\begin{center}
\epsscale{1.15}
\plotone{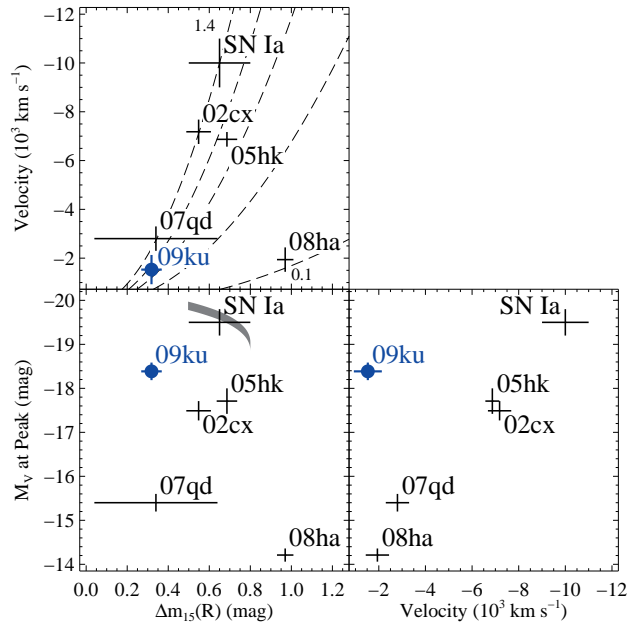}
\caption{A comparison of $\Delta m_{15} (R)$, $M_{V}$ at maximum, and
the photospheric velocity at \about 10~days after maximum (derived
from the minimum of \ion{O}{1} $\lambda 7774$) for various objects in
the SN~2002cx-like class.  The grey band represents the
width-luminosity relationship for SNe~Ia (in the $M_{V}$ vs.\ $\Delta
m_{15} (R)$ plot) and typical values for a Branch-normal SN~Ia.  The
dashed lines represent the relationship between ejecta velocity and
$\Delta m_{15}$ for (from bottom-right to top-left) 0.1, 0.4, 0.7,
1.0, and 1.4~M$_{\sun}$ of ejecta mass.  \sn\ is marked with the blue
circle.}\label{f:prop}
\end{center}
\end{figure}

There is a correlation between decline rate and ejecta velocity;
however, SN~2008ha is an outlier.  Using Arnett's Law
\citep{Arnett82}, we can derive a relationship between ejecta velocity
and $\Delta m_{15} (R)$ (if we assume that the $R$ band is
representative of the light-curve shape of the bolometric light curve)
that depends on ejecta mass.  By assuming that a typical SN~Ia has
1.4~M$_{\sun}$ of ejecta, we obtain curves of constant ejecta mass
which we present in the upper-left panel of Figure~\ref{f:prop}.  All
SNe except for SN~2008ha are consistent with 1.0 -- 1.4~M$_{\sun}$ of
ejecta.

Stretching the \gri\!\! light curves of SN~2005hk to match those of
\sn\ before peak, we derive a rise time of $18.2 \pm 3.0$~days.
Correcting the \gri\! peak magnitudes of \sn\ with the \gri\!
bolometric corrections from SN~2005hk, we find that \sn\ had a peak
bolometric luminosity of $(6.4 \pm 1.6) \times
10^{42}$~ergs~sec$^{-1}$.  Applying Arnett's Law \citep{Arnett82} to
the rise time and peak luminosity of \sn, we find that \sn\ had a
$^{56}$Ni mass of $0.3 \pm 0.1$~M$_{\sun}$.  This is larger than that
found for SN~2005hk (0.22~M$_{\sun}$; \citealt{Phillips07}) but less
than typical SN~Ia explosions (\about 0.4 -- 0.9~M$_{\sun}$; e.g.,
\citealt{Stritzinger06}).

Finally, there does appear to be a relationship between peak
luminosity and ejecta velocity, {\it except for \sn}, which has a much
lower velocity than its peak luminosity suggests (or has a much higher
peak luminosity than its velocity suggests) and is a significant
outlier.


\section{Discussion and Conclusions}\label{s:disc}

PS1 is expected to discover several dozen SN~2002cx-like objects per
year in the Medium-Deep Survey alone (S.\ Rodney, private comm.).
\sn\ is the first of this class to be discovered (and
spectroscopically confirmed) by PS1.  \sn\ is also the most distant
member of the class (400~Mpc; followed by SN~2007qd at 175~Mpc).
Although the early survey PS1 data had a gap in the post-maximum light
curve, measurements from additional telescopes were able to partially
fill this gap.

\sn\ peaked at $M_{V} = -18.4$~mag, which is fainter than typical
SNe~Ia, but also slightly brighter than SNe~2002cx and 2005hk at peak
\citep{Li03:02cx, Phillips07}.  \sn\ declined very slowly in \gri\
relative to other SN~2002cx-like objects, with $\Delta m_{15} (r) =
0.32 \pm 0.05$~mag.  Despite the relatively high peak luminosity and
slow decline, \sn\ had spectra that are most similar to the very
low-luminosity and fast-declining SN~2008ha \citep{Foley09:08ha,
Valenti09, Foley10:08ha}, including a very low ejecta velocity ($v
\lesssim 2000$~\kms).

Although the spectra of SNe~2008ha and \snno\ are similar, \sn\ had a
stronger \ion{Si}{2} $\lambda 6355$ feature and weaker [\ion{Ca}{2}]
$\lambda\lambda 7291$, 7324 and \ion{Ca}{2} NIR triplet features.
Although these differences may be related to ionization or opacity
effects, another possibility is that the burning products of the
explosion produced a different ratio of Si to Ca in the two
explosions.  This could be explained by different degrees of both He
and C/O burning \citep[e.g.,][]{Perets09}, or simply by a different
efficiency for the burning in both explosions.  Disentangling these
effects are beyond the scope of this study, but may provide additional
insight into the explosions of these objects.  Indeed, this may be the
key difference between two very spectroscopically similar SNe that
differ by a factor of \about 40 in peak luminosity.

\citet{McClelland10} suggested that there was a relationship between
light-curve shape, peak luminosity, and ejecta velocity for
SN~2002cx-like objects, including SN~2008ha.  Their sample does show a
relationship between ejecta velocity and peak luminosity; however,
\sn\ is a clear outlier to this relationship.  The class of
SN~2002cx-like objects appear to be quite heterogeneous, with
SNe~2008ha and \snno\ providing examples for the diversity.

With PS1, we expect to discover enough SN~2002cx-like objects to map
the parameter space of the class.  Additionally, by probing parameter
space neglected until recently, PS1 will continue to find rare and
interesting objects.  \sn\ is an example of the diverse transient
phenomena that PS1 will discover.  In the coming years, our
understanding of this class and other exotic transients will be
greatly improved by PS1 and eventually LSST.

\begin{acknowledgments} 

{\it Facilities:}
\facility{PS1(GPC1), Gemini:South(GMOS), Magellan:Clay(MagE), NOT(ALFOSC)}

G.N.\ is supported by NSF award AST--0507475. R.J.F.\ is supported by a Clay
Fellowship. Supernova research at Harvard is supported by NSF grant AST--0907903. 

We are grateful to the staffs at the Gemini, Las Campanas, and Nordic Optical
Telescope Observatories for their dedicated services. We thank E.\ Kankare, S.\
Mattila and S.\ Valenti for their contributions to this work.

The PS1 Surveys have been made possible through contributions of the Institute
for Astronomy at the University of Hawaii in Manoa, the Pan-STARRS Project
Office, the Max-Planck Society and its participating institutes, the Max Planck
Institute for Astronomy, Heidelberg and the Max Planck Institute for
Extraterrestrial Physics, Garching, The Johns Hopkins University, the
University of Durham, the University of Edinburgh, the Queens University
Belfast, the Harvard- Smithsonian Center for Astrophysics, and the Los Cumbres
Observatory Global Telescope Network, Incorporated.

\end{acknowledgments}

\end{document}